# Coexistence of surface oxygen vacancy and interface conducting states in LaAlO$_3$/SrTiO$_3$ revealed by low-angle resonant soft X-ray scattering

*Ming Yang[1,*], Ariando Ariando[2,*], Caozheng Diao[3], James C Lee[4,5], Kaushik Jayaraman[2], Mansoor B A Jalil[6], Serban Smadici[7], Shengwei Zeng[2], Jun Zhou[8], Weilong Kong[2], Mark B. H. Breese[2,3], Sankar Dhar[6], Yuan Ping Feng[2], Peter Abbamonte[5], Thirumalai Venkatesan[2,6], Andrivo Rusydi[2,3,*]*

[1] Department of Applied Physics, The Hong Kong Polytechnic University, Hung Hom, Hong Kong SAR, China

[2] Department of Physics, National University of Singapore, Singapore 117542, Singapore.

[3] Singapore Synchrotron Light Source, National University of Singapore, Singapore 117603, Singapore.

[4] Frederick Seitz Materials Research Laboratory, University of Illinois at Urbana-Champaign Urbana, Illinois 61801, USA.

[5] Brookhaven National Laboratory, Upton, New York 11973, USA.

[6] Department of Electrical and Computer Engineering, National University of Singapore 117576, Singapore.

[7] Dept. of Physics and Astronomy University of Louisville, KY 40292, USA.

[8] Institute of Materials Research and Engineering, Agency for Science, Technology and Research (A*STAR), 2 Fusionopolis Way Singapore 138634, Singapore

**Correspondence to who should be addressed:** M.Y (mingyang@polyu.edu.hk), A. A. (phyarian@nus.edu.sg) or A. R. (andrivo.rusydi@nus.edu.sg)

**Abstract**

Oxide heterostructures have shown rich physics phenomena, particularly in the conjunction of exotic insulator-metal transition (IMT) at the interface between polar insulator $LaAlO_3$ and non-polar insulator $SrTiO_3$ ($LaAlO_3/SrTiO_3$). Polarization catastrophe model has suggested an electronic reconstruction yielding to metallicity at both the interface and surface. Another scenario is the occurrence of surface oxygen vacancy at $LaAlO_3$ (surface-$O_v$), which has predicted surface-to-interface charge transfer yielding metallic interface but insulating surface. To clarify the origin of IMT, one should probe surface-$O_v$ and the associated electronic structures at both the surface and the buried interface simultaneously. Here, using low-angle resonant soft X-ray scattering (LA-RSXS) supported with first-principles calculations, we reveal the co-existence of the surface-$O_v$ state and the interface conducting state *only* in conducting $LaAlO_3/SrTiO_3$ (001) films. *Interestingly*, both the surface-$O_v$ state and the interface conducting state are absent for the insulating film. As a function of $O_v$ density, while the surface-$O_v$ state is responsible for the IMT, the spatial charge distribution is found responsible for a transition from two-dimensional-*like* to three-dimensional-*like* conducting accompanied by spectral weight transfer, revealing the importance of electronic correlation. Our results show the importance of surface-$O_v$ in determining interface properties and provides a new strategy in utilizing LA-RSXS to directly probe the surface and buried interface electronic properties in complex oxide heterostructures.

# 1. Introduction

Since the discovery of an unexpected conducting interface in the atomically abrupt heterostructures comprised of LaAlO$_3$ and SrTiO$_3$ perovskite insulators,[1] many efforts have been devoted to unravelling the origin of this insulator-metal transition (IMT).[1–9] Along with the quasi two-dimensional (2D) conducting interface, various exotic properties that do not exist in their individual bulk counterparts have also been observed at the LaAlO$_3$/SrTiO$_3$ interface, such as Kondo scattering of metallic carriers, interfacial magnetism, and superconductivity.[5,10–19] It is believed that the key source of these fascinating properties is that LaAlO$_3$ is a highly polar insulator, which should result in a diverged polar potential with the increase of LaAlO$_3$ thickness. This leads to an important question – *how does the system overcome such a polar divergence*? The polar catastrophe model suggests that 0.5 electrons transferred from the LaAlO$_3$ surface into the interface can overcome the polar divergence, yielding both metallic interface and surface.[2–4] While this model can explain the observed conducting interface, it *cannot* explain the insulating LaAlO$_3$ surface and other phenomena.[7,9] Therefore, understanding of the surface and how it affects the interface are critical.

It is well-accepted that the emerging IMT property is dependent upon the growth condition. Transport measurements have shown that for samples grown at low oxygen partial pressure ($P_{O2}$ ~$10^{-5}$ and $10^{-6}$ mbar), the interface is highly conducting with a carrier density of ~$10^{17}$ cm$^{-2}$, and the conducting carriers extend deeply into the SrTiO$_3$ substrate.[5,6,14] For samples grown at higher $P_{O2}$ (~ $10^{-3}$ and $10^{-4}$ mbar), on the other hand, the interface has a much lower carrier density of 1~2×$10^{13}$ cm$^{-2}$ with the carriers confined at the interface.[5,6,19] These observations were further confirmed by the atomic force microscopy measurement (AFM), as it revealed a broad carrier distribution for the samples prepared at low $P_{O2}$ and a much narrower distribution for the samples prepared at high $P_{O2}$.[20] It has been reported that the varied $P_{O2}$ could break the interface stoichiometry of LaAlO$_3$/SrTiO$_3$, which might affect interface properties.[8,19] These results have indicated important roles of defects in the IMT and interfacial conductivity, and previous studies have connected them to the effect of surface $O_v$.[21,22] First-principles calculations further highlighted that the occurrence of surface-$O_v$ is the key for the 2D conducting interface and insulating surface,[7,9,19,23,24] which however, is yet to be directly confirmed with experimental evidence.

It is worth mentioning that transport measurements only probe partially of the total charge at the interface, lacking the information about the charge depth profile.[3,4] While other experimental techniques such as photoelectron spectroscopies[25,26] and transmission electron microscopy[22] had been applied but, as they are surface sensitive technique, they may alter the sample surface unintentionally, which in turn, influence its electronic structures. It has been shown that under an intense X-ray irradiation, the 2D conductivity in the $LaAlO_3/SrTiO_3$ might be changed to 3D.[27] Therefore, an alternate, non-destructive experimental method is needed to unravel the surface-$O_v$ and probe electronic structure from the surface to interface for these perovskite oxide heterostructures prepared under different growth conditions. Such measurement should reveal the underlying mechanism that results in IMT and enables us to understand the influence of preparation ambience on electronic properties of the system. In this article, by using low-angle resonant soft X-ray scattering (LA-RSXS), we report direct evidence of both surface-Ov state and conducting interface state in two-dimensional-*like* conducting $LaAlO_3/SrTiO_3$ interface. Supported by first-principles calculations, we further correlate the varied surface-$O_v$ density with the different spatial distribution of interfacial charge changing from insulating, 2D-*like* and to three-dimensional (3D)-*like* conducting (together with $O_v$ in the $SrTiO_3$) $LaAlO_3/SrTiO_3$ interface grown at various $P_{O2}$.

## 2. Results and discussion

**Fabrication and characterization of LaAlO$_3$/SrTiO$_3$ heterostructures.** As a model system, we design three LaAlO$_3$/SrTiO$_3$ heterostructures with different sheet resistances ($R_s$), at the $P_{O2}$ of 1×10$^{-1}$ mbar (sample S1, as a representative for *insulating*), 1×10$^{-3}$ mbar (sample S2, *2D-like* conductivity), and 1×10$^{-6}$ mbar (sample S3, *3D-like* conductivity), respectively. These samples are grown by depositing a single crystal LaAlO$_3$ target on a TiO$_2$-terminated SrTiO$_3$ (001) substrate using pulsed laser deposition (see the details in Method). The thickness of LaAlO$_3$ films is chosen to be 26 unit cells (~9.9 nm) to ensure sufficient photon intensity for scattering and absorption and therefore we can probe the charge distribution from surface to interface. During the deposition, the growth mode was monitored using *in-situ* reflection high energy electron diffraction (RHEED). The observed RHEED oscillation (see **Figure 1a** indicates a layer-by-layer growth for all the samples grown at the different $P_{O2}$, in line with previous reports.[5,11] After deposition, all samples were cooled down to room temperature (RT) at the same $P_{O2}$ during deposition. The surface morphology of the deposited films is studied using AFM. The samples S2 and S3 show similar surface roughness, as shown in **Figure 1b**, while the surface roughness of sample S1 slightly increases. X-ray diffraction (XRD) shows that the three samples have the same crystalline structure, as their main characteristic peaks are overlap (**Figure 1c**). The minor shift in the sub-peaks at 48.3º might be related to varied surface structures of these three heterostructures, which will be discussed in detail later.

**Transport and LA-RSXS measurement on LaAlO$_3$/SrTiO$_3$ heterostructures.** The $R_s$, carrier density $n$ and Hall mobility $\mu$ of these three model samples are measured using a Van-der-Pauw geometry. The temperature-dependent $R_s$ of the low pressure grown sample S3 in **Figure 1d** shows 1 mΩ/□ at 5 $K$ and increased to about 2.1 Ω/□ at RT, with a nearly constant carrier density of ~5.0×10$^{16}$ cm$^{-2}$ (see **Figure 1e**). The Hall mobility is high to ~1.86×10$^4$ cm$^2$ V$^{-1}$ s$^{-1}$ at 5 $K$, as shown in **Figure 1f**. This agrees with a *3D*-like conductivity as suggested by previous studies.[5,6] When the sample grown at a higher $P_{O2}$ (sample S2), the $R_s$ is significantly increased (see the blue curve in **Figure 1d**). It is about 14 kΩ/□ at RT and decreases to around 1 kΩ/□ at 5 $K$. Correspondingly, the associated carrier density varies slightly between $2.5 \times 10^{13}$ cm$^{-2}$ and $2.9 \times 10^{13}$ cm$^{-2}$, as shown in **Figure 1e**. Compared with sample S3, the Hall mobility in Sample S2 is about one-order of magnitude smaller at 5 $K$ (see **Figure 1f**). This temperature-dependent conductivity is quasi-*2D* and has been ascribed to the coexistence of conducting electrons and localized electrons at the interface.[5,12] When the growth pressure is further increased (sample S1), the $R_s$ increases beyond the measurement limit, showing an insulating behaviour.

The transport data suggests a transition of the electronic phase from an insulating, 2D conducting to 3D conducting state in these three representative samples, which are controlled by the $P_{O2}$ during the growth. To further probe the electronic structures from the surface to the interface of these oxide heterostructures, we carried out non-destructive, element specific LA-RSXS measurements at O $K$ (O 1$s$ → 2$p$) and Ti $L_{3,2}$ (Ti $2p_{3/2,1/2}$ → 3$d$) edges. The LA-RSXS experiment is especially designed that the incoming photons is directed normal to the sample surface, while the angle of outgoing photon ($\theta_{out}$) is varied from ~1.0° to ~20.0° with respect to the sample surface (see the illustration in **Figure 2a**). To validate our method, we first examine the LA-RSXS at Ti $L_{3,2}$ edges of the insulating sample S1. Due to the X-ray selection rule, Ti $L_{3,2}$ edges probe the transition from Ti 2$p$ to Ti 3$d$, therefore these edges are sensitive to the presence of Ti ions. As **Figure 2b** shows, we only observe a background without any structure for the $\theta_{out}$ up to 5.0° (see also **Figure S1**, Supporting Information). This reveals that the signals are mainly from the LaAlO$_3$ layers. Interestingly, the features at Ti $L_{3,2}$ edges can be resolved (see **Figure 2b**) when the $\theta_{out}$ is further increased by only 0.5° (from 5.0° to 5.5°), which indicates a sharp electronic structure at the interface of the sample.

The photon penetration depth, $L_p$ at very low angles (< ~10⁰) is governed by the following equation [28]:

$$L_p = \frac{\lambda}{4\pi} \frac{1}{Im(\sqrt{\theta^2 - 2\delta - 2iK})} \quad (1)$$

where $\delta$ and $K$ come from the definition of complex refractive index $N \equiv (1 - \delta) + i K$. For small angles, when θ increases, there is a rapid increase in the penetration depth. The photon penetration depth obtained from the X-ray tabulated value[29] for LaAlO$_3$ at 453.5 eV, which is just below the Ti $L_3$ main edge of STO, is presented in **Figure S3-1** (Supporting Information). It shows that the value at $\theta_{out}$ ~5⁰ is ~ 10.6±1 nm whereas at ~5.5⁰, it is ~14.6±1 nm. Since the LAO film thickness is 10±1nm, therefore, when $\theta_{out}$ is ~5.5⁰, the photon starts probing Ti at the buried interface of LAO/STO and STO. This is consistent with LA-RSXS where Ti $L_{3,2}$ edges are now visible at $\theta_{out}$ ~ 5.5⁰ as shown in **Figure 2b**.

**Observation of surface oxygen vacancy and interface conducting states.** Our main observations are the LA-RSXS results at O *K* edge for samples S1, S2, and S3 as shown in **Figure 3** and **Figure 2c-d**. **Figure S3-2** (Supporting Information) shows the LA-RSXS profiles of all three samples at different $\theta_{out}$. The LA-RSXS is measured in total fluorescence yield (TFY) mode, which is a photon-in-photon-out measurement method which depends on $\theta_{out}$. The data at 2.5⁰ and 20⁰ are shown in **Figure 3a-c**. The experimental data is scaled to the tabulated cross-section obtained from the X-ray tabulated value.[29,30] This gives us the extinction coefficient,[30] *K* of the samples as shown in **Figure S3-2** (Supporting Information). From the penetration depth calculations, we find that when the $\theta_{out}$ < ~ 4.1°, the resonant peaks are mainly from the O ions in the LaAlO$_3$ films (from the LaAlO$_3$ surface to the LaAlO$_3$/SrTiO$_3$ interface). Whereas, for $\theta_{out}$ above ~4.1°, the peaks are contributed by the O from both the LaAlO$_3$ and SrTiO$_3$. For $\theta_{out}$ ~ 2.5°, the photon penetration depth is estimated to be 3.7 ± 1.0 nm, therefore it reveals electronic structure near LaAlO$_3$ surface. For the insulating sample S1, we can see only a dominant peak at ~536.0 eV (**Figure 3a**), which is related to transitions from O 1$s$ to hybridized O 2$p$ / La 5$d$ orbital at LaAlO$_3$ (see the PDOS in **Figure S4**, Supporting Information).[31] For a larger $\theta_{out}$ ~ 20.0°, the photon penetration depth is estimated to be above 50 nm, thus it can detect the signals from both the LaAlO$_3$ films (~10 nm) and the SrTiO$_3$ substrate. A strong peak is observed at ~530.5 eV for the insulating sample S1, while the peak at 536.0 eV remains (the lower panel in **Figure 3a**). The peak at 530.5 eV is ascribed to transitions from O 1$s$ to hybridized O 2$p$ / Ti 3$d$ orbitals in the SrTiO$_3$ near the interface (see the PDOS in **Figure S5**, Supporting Information).

Interestingly, at the small angle ($\theta_{out}$ ~ 2.5°) we find a new pre-peak for both 2D and 3D conducting samples. For the 2D conducting sample S2, the pre-peak is noticeable at ~530.5 eV. We would like to emphasize that the *effective* photon penetration depth at $\theta_{out}$ ~ 2.5° and ~530.5 eV is ~3.7 nm. Thus, due to X-ray dipole allowed transition, this pre-peak shall have O 2$p$ character and is mainly coming from the LaAlO$_3$ surface. Indeed, our theoretical calculations confirm that the pre-peak is a surface O$_v$ state with O 2$p$ character as discussed later. The intensity of this pre-peak grows rapidly as $\theta_{out}$ increases, and it is eventually resonant with the peak from SrTiO$_3$ as seen at higher $\theta_{out}$ (see the lower panels in **Figure 3b**). It is also found that the pre-peak is absent in the insulating sample and the occurrence of it in the conducting samples is accompanied by spectral weight transfer from higher energy structures, particularly 532.4 eV. Such a spectral weight transfer reveals the importance of electronic correlations resulting conducting interface and very different surface electronic structures between the insulating and conducting samples. Intriguingly, by comparing the conducting S2 and S3 samples (**Figures 3b-c**), the pre-peak in the 3D conducting sample is more pronounced than that of the 2D conducting sample. Based on the growth condition, sample S3 is expected to have a higher density of O$_v$, while samples S2 should have a lower density of O$_v$. Thus, the pre-peak in upper panels of **Figure 3b** and **3c** is a characteristic feature of O$_v$ near the LaAlO$_3$ surface in the conducing LaAlO$_3$/SrTiO$_3$ samples. Further details of the pre-peak are given in **Figure S3-4** (Supporting Information).

**Figure 2c** shows the extinction coefficient $K$ of the three samples as a function of $\theta_{out}$. It shows the $K$ taken at three different photon energies: an off-resonance of ~527 eV, pre-peak of ~530.5 eV and the main O $K$-edge of ~536 eV. From the LA-RSXS in **Figure S3-2**, it is clear that the $K$ is more or less the same for all sample for all $\theta_{out}$ when measured at ~527 eV (off-resonance) and ~536 eV (main O $K$ edge). This is clearly seen in **Figure 2c** where $K$ is similar for all three samples at these two energies. Interestingly, we see significant changes at the pre-peak ~530.5 eV, which is the centre of the discussion. The LA-RSXS is an element specific technique and due to X-ray dipole allowed transition, La-RSXS at O $K$ edge is O 1$s$ → 2$p$ transition, therefore it is directly probe unoccupied O 2$p$. As further discussed below and in the Supporting Information for **Figure S3-3**, the photon penetration at $\theta_{out}$ ~ 2.5º is still in the initial few layers of LaAlO$_3$. For LA-RSXS at this grazing angle, the observed pre-peak shall have O 2$p$ character and it is directly probing the surface of LaAlO$_3$. Indeed, our theoretical calculations confirm that the pre-peak is a surface O$_v$ state with O 2$p$ state as discussed below. Intriguingly, the pre-peak is only observed in the conducting samples, which contain surface O$_v$ state, while no pre-peak in the insulating sample. This is further confirmed by the fact that the pre-peak observed at smaller $\theta_{out}$ becomes more pronounced for the samples that are prepared at lower oxygen partial pressure.

We next highlight that in LA-RSXS as a function of $\theta_{out}$ we are probing two different charges, i.e. (1) localized-*like* charges at and near the surface of LAO, which are due to the existence of the surface-*like* oxygen vacancies (surface-O$_V$), and (2) interface-*like conducting* charges at the LaAlO$_3$/SrTiO$_3$ interface, which are induced by a charge transfer due to the presence of O$_V$ at the surface of LaAlO$_3$. We demonstrate the powerfulness of LA-RSXS to probe and distinguish these surface-*like* and interface-*like* charges in the complex oxide heterostructures. The LA-RSXS at a particular $\theta_{out}$ (shown in **Figure 2c**) means an accumulation of the *K* from each layer (or unit cell) till that particular angle and hence a particular depth. The *K* near the interface of LaAlO$_3$ and SrTiO$_3$ has a contribution from both the surface-O$_V$ as well as the charge at the interface.

The *K* seen in **Figure 2c** at any particular angle, and hence a particular depth, is an accumulation of contribution from all the layers above that. To put it in simple terms, we can liken this to an integration of the layer-by-layer contribution till that depth. To get an idea regarding a layer-by-layer distribution, we take the first derivative of the quantity measured in **Figure 2c** to obtain **Figure 2d**, named $\Delta K_{charge}$. The dielectric susceptibility $\chi(\omega, \theta)$, is related to the extinction coefficient through the relation $\sqrt{1 + \chi(\omega, \theta)} = (1 - \delta(\omega, \theta)) + i K(\omega, \theta)$. Scattering of X-rays from charges at a particular energy, gives us information about the charge distribution and about the dielectric susceptibility $\chi$ at that energy.[30,32] So, we argue that the variation of *K* is related to the spatial variation of dielectric susceptibility $\chi(z)$. In our case, the derivative $\Delta K_{charge}$ is qualitatively proportional to the charge density and gives us an idea about the nature of spatial variation of the same. For further analysis, the $\Delta K_{charge}$ is fitted with a Lorentzian fitting resulting a full width at half maximum (FWHM) of ~ 1.8º and ~ 5.9º for the S2 sample and S3 sample, respectively. This reveals that the S2 sample has the charge distribution of 2D-*like* and is mostly concentrated and high in the SrTiO$_3$ near the LaAlO$_3$/SrTiO$_3$ interface, whereas in the case of the S3 sample, it is distributed 3D-*like* over LaAlO$_3$ to SrTiO$_3$ crossed the interface. Similar charge carrier distributions in the conducting LaAlO$_3$/SrTiO$_3$ heterostructures have been mapped previously using CT-AFM.[20] For further details, the penetration depth as a function of $\theta_{out}$ is plotted and discussed in **Figure S3-3** of Supporting Information.

**First-principles calculations.** To find the origin of those pre-peaks and its impact on the insulating surface and conducting interface, we perform first-principles calculations using four idealized models: (2×2×1) 5 layers of LaAlO$_3$ supercell on the SrTiO$_3$ (001) substrate without surface-O$_v$ (as a representative of insulating), with one surface-O$_v$ (**Figure 4a**) (as representative of 2D-*like* conducting), with two surface-O$_v$ (3D conducting) and with one surface-O$_v$ and one interface-O$_v$ (as representative of 3D-like conducting). From the layer-projected density of states (PDOS) for the LaAlO$_3$/SrTiO$_3$ interface without O$_v$ (see **Figure S6**, Supporting Information), there are no mid-gap states observed. This corresponds with the LA-RSXS for the insulating sample S1 where no pre-peak is observed at small $\theta_{out}$. We further examine LaAlO$_3$/SrTiO$_3$ interface models with other intrinsic defects, including cation vacancies (La, Sr, Ti and Al vacancy), Al-Ti antisite defect, and Sr-La antisite defect. However, due to their high formation energy (see **Figures S7-9**) and the absence of mid-gap states resulting from the surface LaAlO$_3$ layer (as shown **Figure S8**), it is unlikely that these defects contribute to the pre-peak in the O $K$ edge observed at small $\theta_{out}$.

In contrast, the presence of O$_v$ at the LaAlO$_3$ surface introduces a new mid-gap state (**Figures 4b** and **4c** and **Figures S10-S12** in Supporting Information). The PDOS also shows that such the surface of LaAlO$_3$ remains insulating due to a charge transfer from the surface to the interface. This mid-gap state is mainly ascribed to the *p* orbitals of the surface oxygen atoms, as well as weaker contribution from surface Al atoms. It is mainly from the surface layer as evidenced by the visualized partial charge density (blue colour in **Figure 4a**). The mid-gap state induced by the surface-O$_v$ is resonant with the interfacial conducting states (**Figure 4c**) from the interfacial Ti t$_{2g}$ orbitals. This can be understood as the excess charges induced by surface-O$_v$ are transferred into the interfacial TiO$_2$ layers to minimize the polar divergence in the LaAlO$_3$ layers.[7,9] Furthermore, in the case of LaAlO$_3$/SrTiO$_3$ heterostructure with higher density of O$_v$, for example, two surface-O$_v$ in the model structure, it turns out that the mid-gap state becomes slightly more pronounced (**Figure 4d** and **Figures S13-S14** in Supporting Information).

It has been suggested that when the LaAlO$_3$/SrTiO$_3$ heterostructure is prepared at the lower $P_{O2}$, the O$_v$ might occur in the SrTiO$_3$ substrate.[6,23] Indeed, from calculated layer dependent formation energy of O$_v$ (**Figure S9** in Supporting Information), the formation of O$_v$ at the SrTiO$_3$ interface is possible as its formation energy is the second lowest (the lowest one is for both O$_v$ at the LaAlO$_3$ surface as we discussed above) for the oxygen deficient growth condition. This interfacial O$_v$ introduces additional charge carriers, which extend much deeper into the SrTiO$_3$ substrate (**Figure S15**, Supporting Information), thus lending to more pronounced 3D conductivity, compared with 2D conducting and insulating samples (see the comparison in **Figure S16**). These calculation results collaborate well with the experimental observations, which clearly suggest the existence of surface O$_v$ in the conducting LaAlO$_3$/SrTiO$_3$ interface and unravel surface effects on the interface properties in the 2D-*like* conducting sample. While for the 3D-*like* conducting sample, the coexistence of the O$_v$ near LaAlO$_3$ surface and the O$_v$ in SrTiO$_3$ near the interface contributes to the more extended conducting carrier in the samples, in which the interfacial O$_v$ introduces additional electronic states below Fermi level that can be decoupled with that of surface O$_v$ (see **Figures S15**).

It is believed that the conducting interface of the LaAlO$_3$/SrTiO$_3$ heterostructure is due to the charge transfer from the LaAlO$_3$ layers into the Ti ions at the interface.[2,7,9,19] Our detailed LA-RSXS measurements supported with first-principles calculations show that surface-O$_v$ occurs yielding this charge transfer process and plays a critical role in the onset of interface conductivity. Without surface-O$_v$, charge transfer is very weak, if any, thus resulting in an insulting behaviour as evidenced by the sample prepared at high $P_{O2}$ (S1 sample). In this case, the $O_v$ is either difficult to be formed or can be compensated by the high $P_{O2}$. In contrast, when the samples are prepared at a low $P_{O2}$ (S2 sample), the formation of surface-$O_v$ is favourable. Thus, the excess charges from the surface are transferred into the interfacial Ti ions, leading to the quasi-2D conducting interface while the surface remains insulating. When the samples are prepared at very low $P_{O2}$ (S3 sample), more surface-$O_v$ are formed, and the $O_v$ might occur at the interface or even in the SrTiO$_3$ substrate.[6,33] These lead to a 3D-*like* conductivity and a very broad distribution of carrier density. From our theoretical calculations, the conducting LaAlO$_3$/SrTiO$_3$ (001) has two different electronic structures: (1) at the LaAlO$_3$ surface, the insulating mid-gap state occurs due to surface-$O_v$ and (2) at the LaAlO$_3$/SrTiO$_3$ interface the conducting interface state appears due to the charge transfer into Ti 3*d* orbital. The LA-RSXS reveals these two new states. When the detection angle $\theta_{out}$ is small, LA-RSXS directly probes the surface-$O_v$ states at O *K* edge. When the $\theta_{out}$ is larger, the conducting interface state can be resolved at O *K* and Ti *L$_{3,2}$* edges. Thus, by using LA-RSXS, we can probe spatial distribution electronic structure of LaAlO$_3$ surface down to the buried SrTiO$_3$ interface in the LaAlO$_3$/SrTiO$_3$. We note that in our LA-RSXS, the presence of surface-$O_V$ state is not influenced by the synchrotron radiation beam evident by the absence of the pre-peak in the insulating sample S1.

## 3. Conclusion

In conclusion, using novel LA-RSXS we present direct experimental evidence of co-existence of LaAlO$_3$ surface-$O_v$ state and conducting interface charge state and the corresponding spatial charge distribution at the LaAlO$_3$/SrTiO$_3$ interface. Supported by first-principles calculations, the interfacial conductivity and the charge distribution are found to be strongly dependent on the surface-$O_v$ accompanied by spectral weight transfer. Our result unravels the important role of surface-$O_v$ and electronic correlations in determining the electronic properties at the perovskite oxide interface and demonstrates a new capability of LA-RSXS in probing spatial distribution electronic structure from a surface to a buried interface of oxide heterostructures.

Materials and Methods

*Sample Growth*: To obtain a single TiO$_2$ terminated surface, SrTiO$_3$ are treated by buffered-HF and annealed at 950°C in oxygen environment.[34] With this treatment, an atomically flat SrTiO$_3$ surface is obtained and confirmed by clear unit cell height steps observed with atomic force microscopy.[5] Twenty-six unit cells of LaAlO$_3$ are deposited by using pulsed laser deposition and a single crystal LaAlO$_3$ target on the TiO$_2$-terminated (001) SrTiO$_3$ substrates in oxygen partial pressures ranging from 10$^{-6}$ to 10$^{-1}$ mbar at 850 °C. The films are grown using a laser energy with a fluency of 3 J/cm$^2$ and repetition rate of 1 Hz. The deposition is monitored by periodic RHEED oscillation, which clearly show the layer-by-layer growth for the LaAlO$_3$ films. After growth, all samples are cooled to room temperature at the same oxygen partial pressure used during the deposition.

*Electrical Measurement*: Van der Pauw geometry is employed to measure the sheet resistance, carrier density and Hall mobility of the three samples S1, S2 and S3. The wire connection for electrical transport measurement is done by Al ultrasonic wire bonding and the measurements were performed in Quantum Design Physical Property Measurement System (PPMS).

*Low-angle Resonant Soft X-ray Scattering (LA-RSXS) Measurements*: The LA-RSXS measurements were performed at the soft X-ray and ultraviolet (SUV) beamline of the Singapore Synchrotron Light Source and at the X1B beamline of National Synchrotron Light Source. [30,32,35] The LA-RSXS experiment was especially designed that the incoming photon was fixed parallel to the normal surface of the samples (with electric field of photon perpendicular to the surface of the sample), while the angle of outgoing photon ($\theta_{out}$) was varied from ~1° to ~20.0° with respect to the surface of the samples. The energy resolution was set to ~0.25 eV.

The results of LA-RSXS at the O $K$ and Ti $L_{3,2}$ edges from the three samples with different outgoing angles are shown in Supporting Information (**Figures S1 and S2**). In order to extract the changes in electronic structure at the LaAlO$_3$/SrTiO$_3$ interface, we used SrTiO$_3$ bulk as reference. All the measurements were done at room temperature.

*DFT Calculations*: All the calculations are performed using density-functional-theory based Vienna ab initio simulation package (VASP5.4.18) with the Perdew–Burke–Ernzerhof (PBE) format exchange-correlation functional and the projector-augmented wave (PAW) pseudopotentials.[36–38] The cut-off energy for the expansion of plane-wave basis is set to 500 eV. The on-site Coulomb interaction is considered with effective U=1.2 and U=11 adopted for *d* and *f* orbitals of Ti and La ions, respectively. The LaAlO$_3$/SrTiO$_3$ interfaces are modeled by placing five unit cells of (2×2×1) LaAlO$_3$ supercell on TiO$_2$-terminated SrTiO$_3$ (001) substrates with a thickness of 5 unit cells, in which we consider four interfaces structures: without O$_v$, with a surface-O$_v$, with two surface-O$_v$, and with one surface-O$_v$ and one O$_v$ at the interface TiO$_2$ sublayer. A vacuum with a thickness of 15 Å is applied normal to LaAlO$_3$ surfaces to minimize artificial Coulomb interaction between neighboring surfaces. 6×6×1 Gamma-point-centered k-point meshes are used for sampling Brillouin zones of the LaAlO$_3$/SrTiO$_3$ interface structures. The electronic convergence is set to $1.0\times10^{-6}$ eV, and the force on each atom is relaxed to be smaller than 0.01 eV/Å with fixed bottom layer of the SrTiO$_3$ substrate. The dipole correction is applied to minimize artificial dipole interactions.[39]

**Supporting Information**

Supporting Information is available from the Wiley Online Library or from the author.

**Acknowledgements**

We acknowledge technical support and discussion with Jason Lim and Xiao Renshaw Wang. This work was supported by the Ministry of Education of Singapore (MOE) AcRF Tier-2 (T2EP50220-0041, T2EP50122-0028 and MOE2018-T2-2-117), NRF-NUS Resilience and Growth Postdoctoral Fellowships (R-144-000-455-281 and R-144-000-459-281), and NUS Core Support (C-380-003-003-001). M. Y. would like to acknowledge the funding support from Hong Kong Polytechnic University (Project No.: 1-BE47 and ZE2F). We acknowledge Centre for Advanced 2D Materials and Graphene Research at National University of Singapore, and National Supercomputing Centre of Singapore for providing computing resource. The authors would also like to acknowledge the Singapore Synchrotron Light Source (SSLS) for providing the facilities necessary for conducting the research. The SSLS is a National Research Infrastructure under the National Research Foundation Singapore.

**Conflict of Interest**

The authors declare no conflict of interest.

**Author Contributions**: A.R. designed the LA-RSXS for detecting the electronic structure at surface and interface. A.R., C.D., J.C.L., S.S. performed low-angle resonant soft X-ray scattering. A.A., S.W.Z, and T.V. prepared samples and perform transport and X-ray diffraction and RHEED measurements. M.Y., J.Z., M.J., W.K., A.R., Y.P.F. performed theoretical calculations. M.Y., K.J. and A.R. analyzed data comprehensively and wrote paper with inputs from all co-authors. A.R. and A.A. initiated and led the project.

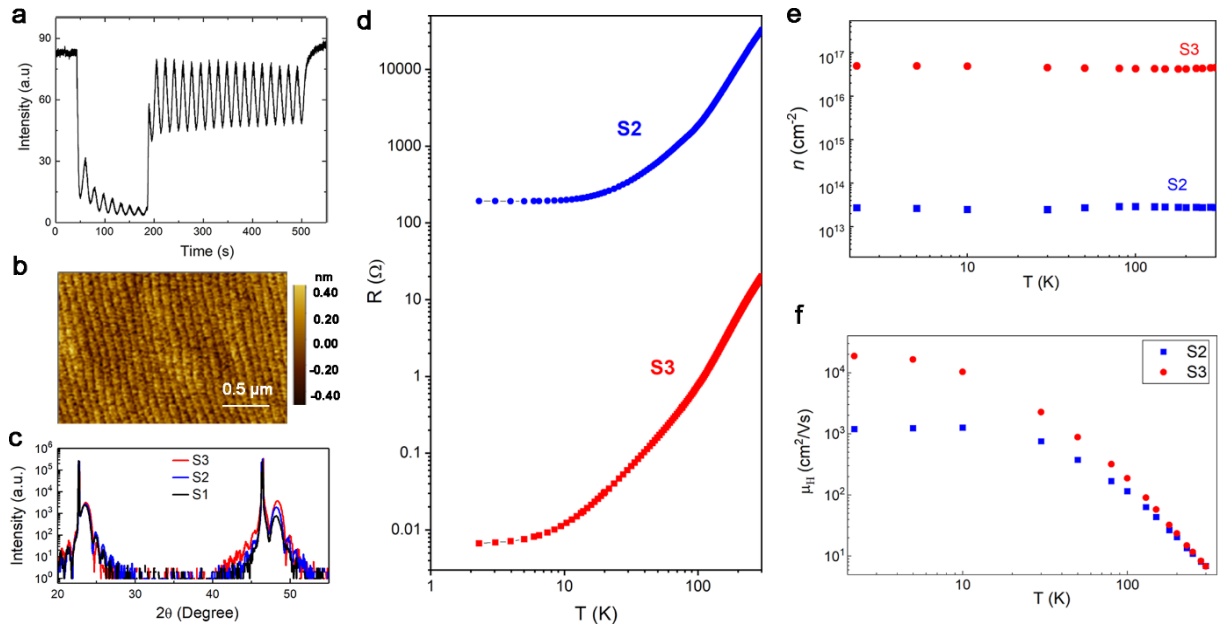

**Figure 1.** Structural and electric characterization of three LaAlO$_3$/SrTiO$_3$ (001) samples (S1-S3) grown at different oxygen partial pressures. (a) Reflection high-energy electron diffraction (RHEED) oscillations during deposition corresponding to 26 unit cells of LaAlO$_3$ on the SrTiO$_3$ (001) substrate terminated with TiO$_2$ layer. (b) The AFM image for sample S2. (c) X-ray diffraction (XRD) (log plot) for the samples S1-S3. (d-f) temperature dependent sheet resistance, carrier density and Hall mobility for sample S2 and S3.

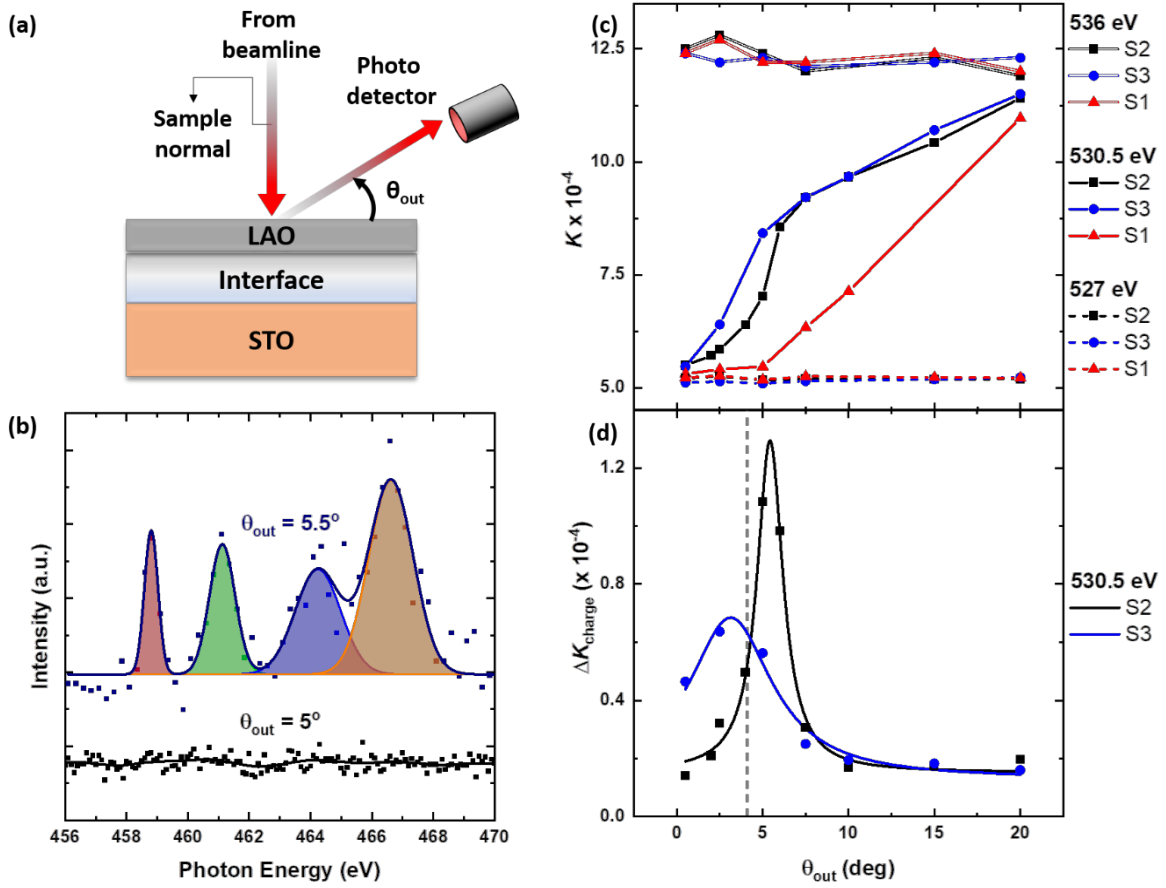

**Figure 2.** (a) Schematic diagram for the LA-RSXS measurements. (b) The LA-RSXS at Ti $L_{3,2}$ edges with the $\theta_{out}$ of 5.0º and 5.5º for sample S2. (c) The plot of extinction coefficient, $K$ of all the samples at off-resonance (~527 eV), the pre-peak (~530.5 eV) and the main O $2p$ peak (~536 eV), respectively. (d) The first-derivative plot of $K$ shown in (c), named $\Delta K_{charge}$, at the pre-peak energy for the S2 (2D-*like*) and S3 (3D-*like*) samples, which represents the distribution of charge. The dashed line in **Figure 2d** around ~4.1º corresponds to a depth of ~10 nm, which is the thickness of the LAO layer.

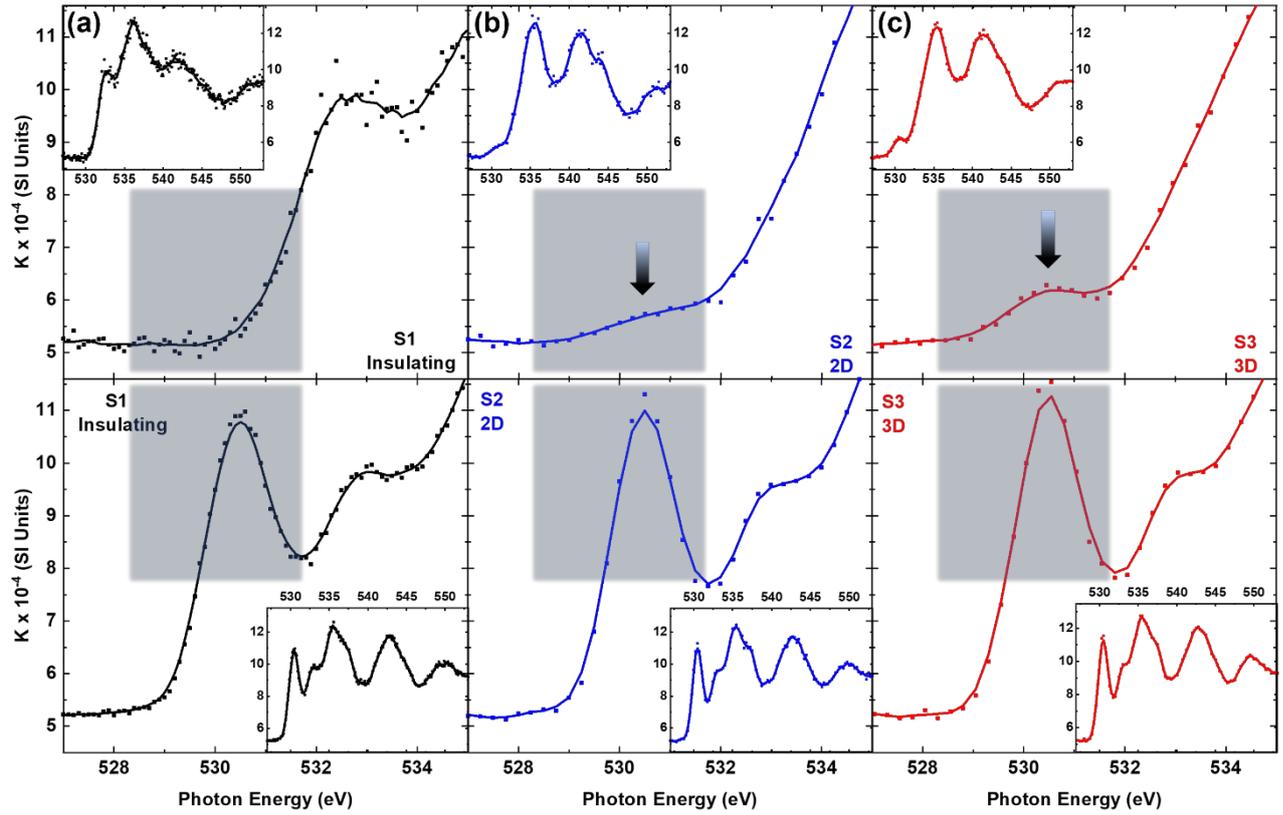

**Figure 3**. The LA-RSXSs at O $K$ edge for sample (a) S1, (b) S2 and (c) S3 at the $\theta_{out}$ of 2.5º (top panel) and 20º (bottom panel). The new pre-peaks are highlighted. The insets show a wider energy range.

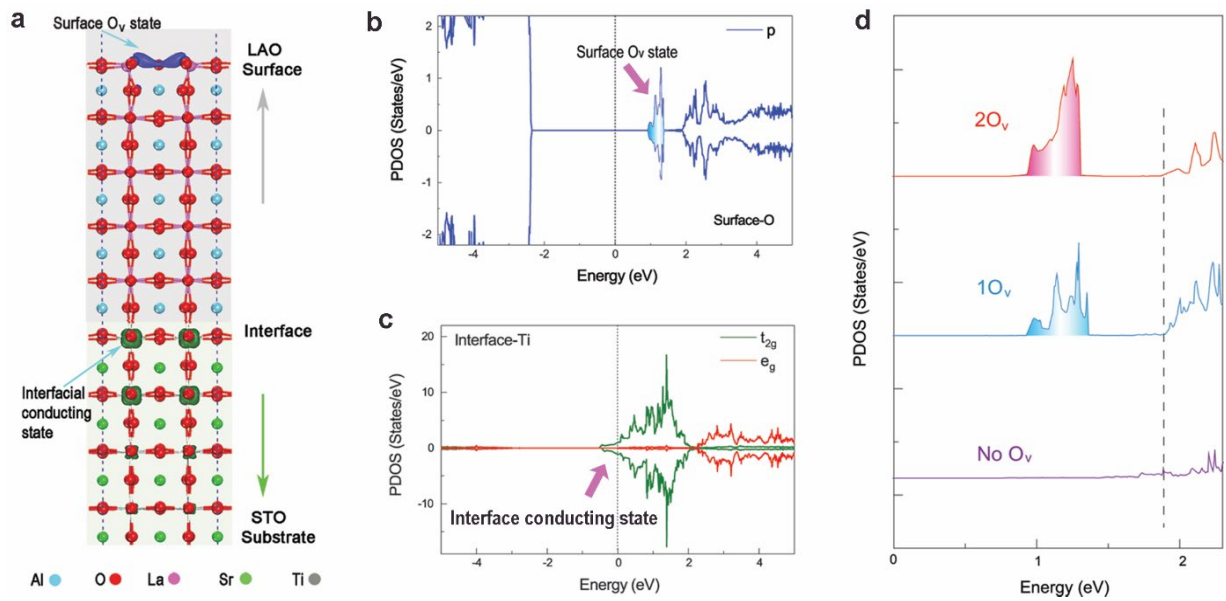

**Figure 4.** (a) An illustration of the atomistic structure of the LaAlO$_3$/SrTiO$_3$ heterostructure with a surface O$_v$, superimposed with the partial charge density of the surface O$_v$ state (blue color) and interfacial Ti T$_{2g}$ states (green color) using an iso-surface value of $1.0\times10^{-3}$ e/Å$^3$. The PDOSs of (b) surface O and (c) interfacial Ti atoms, respectively, where the Fermi level is set to 0 eV. (d) The PDOSs (spin-up) of oxygen atoms at the surface AlO$_2$ layers for the LaAlO$_3$/SrTiO$_3$ heterostructures without the O$_v$, with one surface O$_v$, and with two surface O$_v$, where the vacuum level was used for the alignment.